\input{aipcheck}

\documentclass[
    ,final            
  ,numberedheadings 
  ]
  {aipproc}

\layoutstyle{6x9}
\usepackage{epstopdf}
\usepackage{graphicx}
\def\ket#1{|#1\rangle}
\def\bra#1{\langle#1|}
\def\bz{\beta_0}

\newcommand{\ba}{\begin{eqnarray}}
\newcommand{\ea}{\end{eqnarray}}
\newcommand{\ban}{\begin{eqnarray*}}
\newcommand{\ean}{\end{eqnarray*}}
\newcommand{\bsub}{\begin{subequations}}
\newcommand{\esub}{\end{subequations}}

\begin{document}

\title{Interplay of order and chaos across a first-order quantum 
shape-phase transition in nuclei} 


\author{A. Leviatan}
{address={Racah Institute of Physics, The Hebrew University, 
Jerusalem 91904, Israel}}

\author{M. Macek}
{address={Racah Institute of Physics, The Hebrew University, 
Jerusalem 91904, Israel}}

\begin{abstract}
We study the nature of the dynamics in a first-order
quantum phase transition between spherical and prolate-deformed nuclear 
shapes. Classical and quantum analyses reveal a change in the system 
from a chaotic H\'enon-Heiles behavior on the spherical side into a 
pronounced regular dynamics on the deformed side. 
Both order and chaos persist in the coexistence region and their 
interplay reflects the Landau 
potential landscape and the impact of collective rotations.
\end{abstract}

\maketitle


Shape-Phase transitions in nuclei are an example of 
quantum phase transitions (QPTs) in a mesoscopic (finite) system. 
These are qualitative changes in the properties of the system 
induced by a variation of parameters $\lambda$ in the 
quantum Hamiltonian $\hat{H}(\lambda)$~\cite{ref:Hert76}. 
Such ground-state transformations have become a topic of great interest 
in different branches of physics~\cite{carr}. 
The competing interactions 
that drive these transitions, can affect dramatically the nature of the 
dynamics and, in some cases, lead 
to an intricate interplay of order and chaos. In the present contribution 
we study this effect~\cite{ref:MacLev11,ref:LevMac12} 
in relation to first-order QPTs between spherical 
and axially-deformed nuclei~\cite{ref:Cejn10}, 
as encountered in the Nd-Sm-Gd region.
We employ the interacting boson model (IBM)~\cite{ref:Iac87} 
involving $N$ ($s,d$) bosons 
with angular momentum $L=0,2$, representing valence nucleon pairs.
The model has been widely used in describing QPTs~\cite{ref:Diep80} 
and chaos~\cite{ref:Whel93} in nuclei.

In studying QPTs, it is convenient to resolve 
the IBM Hamiltonian into two parts, 
$\hat{H} = \hat{H}_{\mathrm{int}} 
+ \hat{H}_{\mathrm{col}}$~\cite{ref:Lev87}. 
The intrinsic part ($\hat{H}_{\mathrm{int}}$) 
determines the potential surface $V(\beta,\gamma)$, 
while the collective part ($\hat{H}_{\mathrm{col}}$)
is composed of kinetic terms which do not affect the shape of 
$V(\beta,\gamma)$. Here $(\beta,\gamma)$ are quadrupole shape parameters 
whose values $(\beta_{\mathrm{eq}},\gamma_{\mathrm{eq}})$ at the global 
minimum of $V(\beta,\gamma)$ 
define the equilibrium shape for a given Hamiltonian. 
Focusing on first-order QPTs between 
stable spherical ($\beta_{\mathrm{eq}}=0$) and prolate-deformed 
($\beta_{\mathrm{eq}}>0$, $\gamma_{\mathrm{eq}}=0$) shapes, 
the intrinsic Hamiltonian reads
\ba
\label{eq:H1}
\hat{H}_\mathrm{int}^{I}(\rho)/\bar{h}_2 &=& 
2(1\!-\! \rho^2\beta_0^{2})\hat{n}_d(\hat{n}_d \!-\! 1)
+\beta_0^2 R^{\dag}_{2}(\rho) \cdot\tilde{R}_{2}(\rho) ~,\\
\label{eq:H2}
\hat{H}_\mathrm{int}^{II}(\xi)/ \bar{h}_2 &=& 
\xi P^{\dag}_0 P_0 + 
P^{\dag}_2 \cdot \tilde{P}_2 ~, 
\ea
where $\overline{h}_2\equiv h_2/ N(N-1)$. 
Here $\hat{n}_d$ is the $d$-boson number operator, 
$R^{\dag}_{2\mu}(\rho) \!=\! \sqrt{2}s^\dag d^\dag_\mu + 
\rho\sqrt{7}(d^\dag d^\dag)^{(2)}_\mu$, 
$P^{\dag}_0 \!=\! 
d^\dag \cdot d^\dag - \beta_0^2 (s^\dag)^2$, 
$P^{\dag}_{2\mu} \!=\! 
\sqrt{2}\beta_0 s^\dag d^\dag_\mu + 
\sqrt{7}(d^\dag d^\dag)^{(2)}_\mu$. 
The parameters that control the QPT 
are $\rho$ and $\xi$, 
with $0\leq\rho\leq \bz^{-1}$ and $\xi\geq 0$, 
while $\beta_0$ is a constant. 
$\hat{H}_\mathrm{int}^{I}(\rho)$ and 
$\hat{H}_\mathrm{int}^{II}(\xi)$ are the intrinsic Hamiltonians in the 
spherical and deformed phases, respectively. 
They coincide at the critical point 
$\rho_c\!=\!\bz^{-1}$ and $\xi_c \!=\!0$: 
$\hat{H}_\mathrm{int}^{I}(\rho_c) \!=\! \hat{H}_\mathrm{int}^{II}(\xi_c)$.

\begin{figure}[!t]
\includegraphics[width=0.5\linewidth]{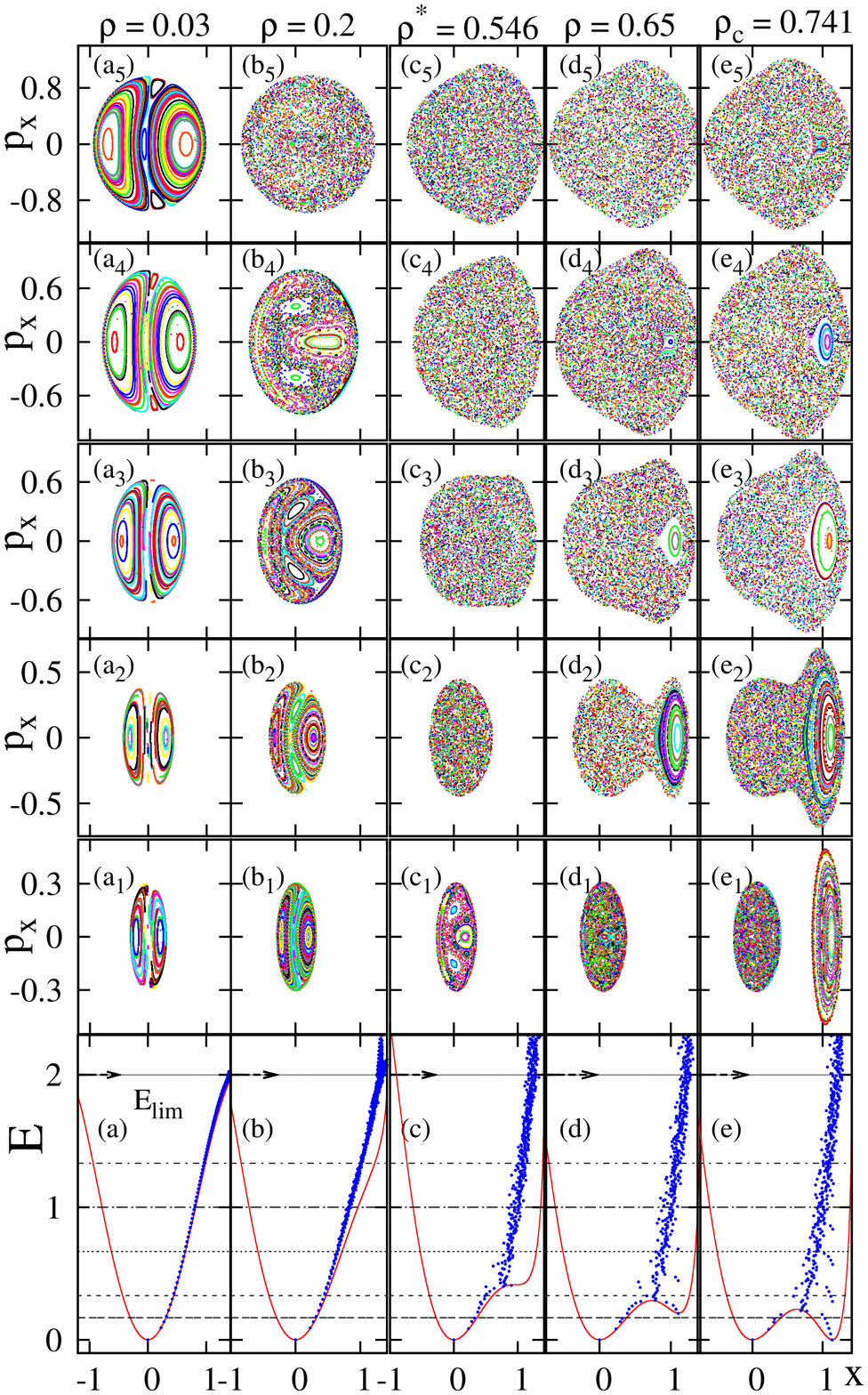}
\hspace{0.1cm}
\includegraphics[width=0.49\linewidth]{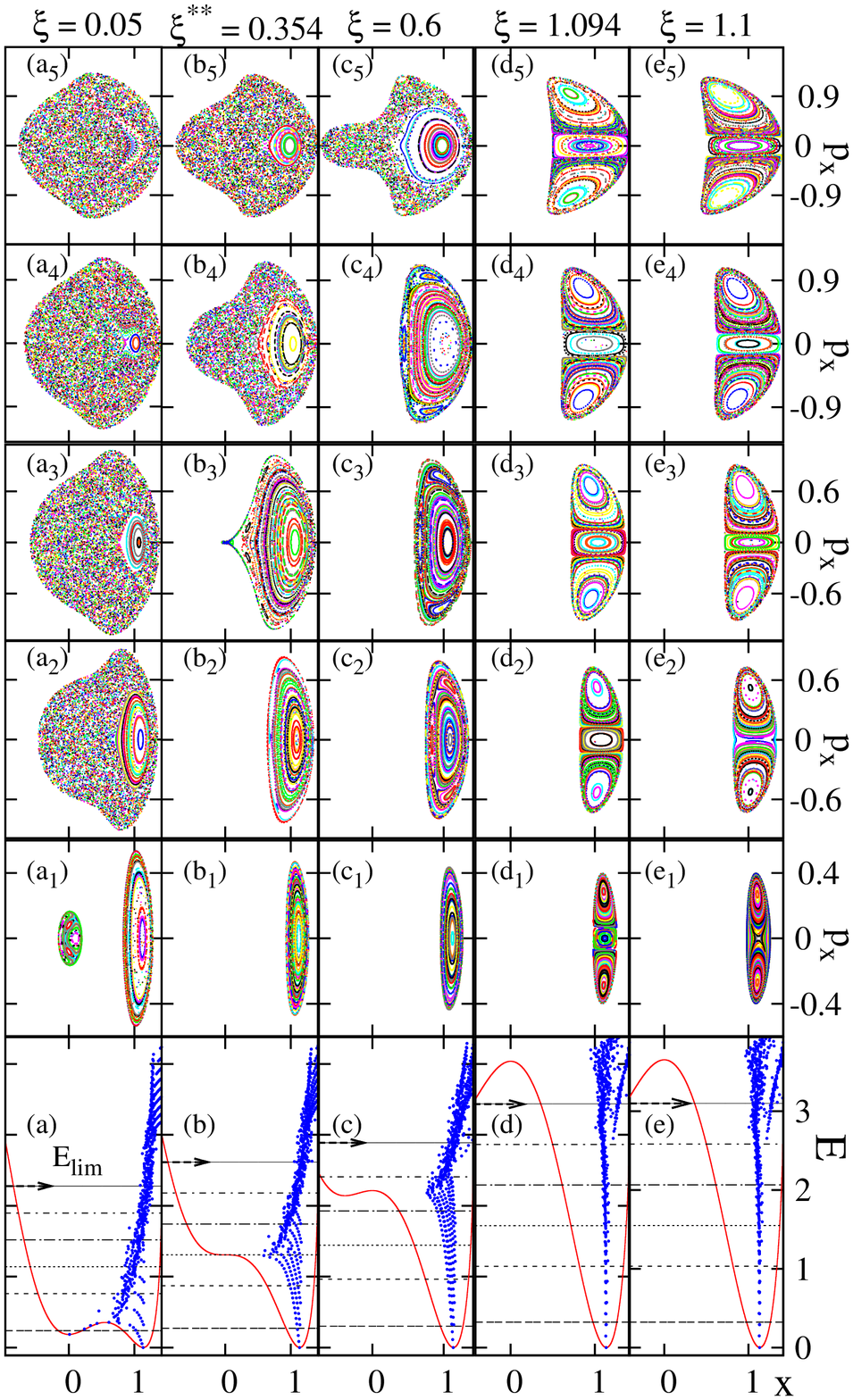}
\caption{
Poincar\'e sections (upper five rows) depicting the classical dynamics 
of $\hat{H}^{I}_{\mathrm{int}}(\rho)$~(\ref{eq:H1}) and 
$\hat{H}^{II}_{\mathrm{int}}(\xi)$~(\ref{eq:H2}) with 
$h_2\!=\!1,\,\beta_0\!=\!1.35$, for several values of 
$\rho\leq\rho_c$ and $\xi>\xi_c$.
The bottom row displays the corresponding classical potentials 
$V^{(I)}(\rho)$~(\ref{eq:V1}) and $V^{(II)}(\xi)$~(\ref{eq:V2}) 
and the five energies, below the domain boundary 
$E_\mathrm{lim}\!=\!V(\beta\!=\!\sqrt{2},\gamma)$, 
at which the sections were calculated. The Peres lattices 
$\{x_i,E_i\}$, portraying the quantum dynamics for eigenstates 
with $L\!=\!0$ and $N\!=\!80$, are overlayed on the 
classical potentials $V(x,y=0)$. 
They exhibit sequences of regular states in the vicinity of the 
deformed well, consisting of 
bandhead states of the ground $g(K=0)$, $\beta^n(K=0)$, 
$\beta^n\gamma^2(K=0)$, $\beta^n\gamma^{4}(K=0)$ 
bands, etc.} 
\label{fig:1}
\end{figure}

The classical limit is obtained through the use of coherent 
states, rescaling and taking $N\rightarrow\infty$, with 
$1/N$ playing the role of $\hbar$~\cite{ref:Whel93}. 
The derived classical Hamiltonian 
involves complicated expressions of shape variables $(\beta,\gamma)$, 
Euler angles and their conjugate momenta. 
Setting the latter to zero, yields the following classical (Landau) 
potentials
\ba
\label{eq:V1}
V^{I}(\rho)/h_2 &=& 
\bz^2 \beta^2 - 
\rho\bz^2 \sqrt{2\!-\!\beta^2} \beta^3\cos3\gamma
+ \textstyle{\frac{1}{2}}(1\!-\!\bz^2)\beta^4~,\\
\label{eq:V2}
V^{II}(\xi)/h_2 &=&  
\bz^2[1 - \xi(1\!+\!\bz^2)] \beta^2
-\bz \sqrt{2\!-\!\beta^2} \beta^3\cos3\gamma
\qquad
\nonumber\\
&&
+ \textstyle{\frac{1}{4}}[2(1\!-\!\bz^2) + \xi(1\!+\!\bz^2)^2]\beta^4
+ \xi\bz^4~.
\quad
\ea
The $(\beta,\gamma)$ variables can be parametrized by Cartesian coordinates 
$x\!=\!\beta\cos{\gamma}$ and $y\!=\!\beta\sin{\gamma}$. 
The potential $V^{I}(\rho)$ [$V^{II}(\xi)$] 
has a global spherical [deformed] minimum with, respectively, 
$\beta_{\mathrm{eq}}\!=\!0$ 
[$\beta_{\mathrm{eq}}\!=\!\sqrt{2}\bz(1+\bz^2)^{-1/2}
\!>\!0,\gamma_{\mathrm{eq}}\!=\!0$]. 
At the spinodal point ($\rho^{*}$), 
$V^{I}(\rho)$ develops an additional local deformed minimum, 
and the two minima cross and become degenerate 
at the critical point $\rho_c$ (or~$\xi_c$). The spherical minimum 
turns local in $V^{II}(\xi)$ for $\xi>\xi_c$ 
and disappears at the anti-spinodal point~($\xi^{**}$). 
The  order parameter ${\beta_{\mathrm eq}}$ is a double-valued function 
in the coexistence region (in-between $\rho^{*}$ and $\xi^{**}$) 
and a step-function outside it. 
The potentials $V(\beta,\gamma=0)\!=\!V(x,y=0)$ 
for several values of $\xi,\rho$, are shown at the bottom row 
of Fig.~\ref{fig:1}. The height of the barrier at the critical point is 
$V_b\!=\!h_{2}[1-(1+\beta_0^2)^{1/2}]^2/2$. 
Henceforth, we set $\beta_0\!=\!1.35$, resulting in a high barrier 
$V_b/h_2 \!=\!0.231$ 
(compared to $V_b/h_2 \!=\!0.0018$ in previous 
works~\cite{ref:Whel93}). 

The classical dynamics of $L\!=\!0$ vibrations, 
governed by $\hat{H}_{\mathrm{int}}$, 
can be depicted conveniently via Poincar\'e sections. 
These are shown in Fig.~\ref{fig:1} for selected energies 
and  control parameters. 
For $\rho\!=\!0$, the system is integrable, with 
$V^{I}(\rho\!=\!0)\!\propto\!\bz^2 \beta^2 + 
\textstyle{\frac{1}{2}}(1\!-\!\bz^2)\beta^4$. 
The sections for $\rho\!=\!0.03$ in Fig.~\ref{fig:1}, 
show the phase space 
portrait typical of an anharmonic (quartic) oscillator 
with two major regular islands, weakly perturbed by the 
small $\rho\cos 3\gamma$ term.
For small~$\beta$, 
$V^{I}(\rho)\!\approx\! \beta^2 \!-\! 
\rho\sqrt{2}\beta^3\cos 3\gamma$. 
The derived phase-space portrait, shown for $\rho\!=\!0.2$ 
in Fig.~\ref{fig:1},  
is similar to the  H\'enon-Heiles system (HH)~\cite{ref:Heno64} 
with regularity at low energy [panels (b$_1$)-(b$_2$)] and 
marked onset of chaos at higher energies [panels (b$_3$)-(b$_5$)]. 
The chaotic component of the dynamics increases with $\rho$ and 
maximizes at the spinodal point $\rho^{*}\!=\!0.546$. 
The dynamics changes profoundly in the coexistence region, 
shown for $\rho\!=\!0.65,\, 0.741$ 
and $\xi\!=\! 0.05$ in Fig.~\ref{fig:1}. 
As the local deformed minimum develops, robustly regular dynamics 
attached to it appears. The trajectories form a single island 
and remain regular at energies well above the barrier height~$V_b$, 
clearly separated from the surrounding chaotic environment. 
As $\xi$ increases, the spherical minimum becomes shallower, 
the HH-like dynamics diminishes and disappears 
at the anti-spinodal point $\xi^{**}\!=\!0.354$. 
Regular motion prevails for $\xi \!>\! \xi^{**}$, where the section 
landscape changes from a single to several regular islands
and the dynamics is sensitive to local degeneracies of 
normal-modes~\cite{ref:LevMac12}.

The quantum manifestations of the rich classical dynamics 
can be studied via Peres lattices $\{x_i,E_i\}$~\cite{ref:Peres84}. 
Here $E_i$ are the energies of eigenstates~$\ket{i}$ of the Hamiltonian 
and $x_i \!\equiv\! \sqrt{2\bra{i}\hat{n}_d\ket{i}/N}$. 
The lattices can distinguish regular from irregular states 
by means of ordered patterns and disordered 
meshes of points, respectively. 
The particular choice of $x_i$ can associate the states with 
a given region in phase space through the classical-quantum correspondence 
$\beta \!=\!x \!\leftrightarrow\! x_i$~\cite{ref:MacLev11}. 
The Peres lattices for $L\!=\!0$ eigenstates of 
$\hat{H}_\mathrm{int}$ with $N\!=\!80$, 
are shown on the bottom row of Fig.~\ref{fig:1}, 
overlayed on the classical potentials $V(x,y=0)$.
For $\rho\!=\!0$, the Hamiltonian~(\ref{eq:H1}) 
has U(5) dynamical symmetry with a solvable spectrum 
$E_i \!=\! 2\bar{h}_2[\bz^2N\!-\!1 + (1\!-\!\bz^2)n_d]n_d$. 
For large $N$, the corresponding Peres lattice coincides 
with $V^{I}(\rho=0)$, a trend seen in Fig.~\ref{fig:1}. 
Whenever a deformed minimum occurs in the potential, the Peres lattices 
exhibit regular sequences of states, localized in the region of 
the deformed well and persisting 
to energies well above the barrier. 
They are related to the regular islands in the Poincar\'e sections 
and are well separated from the remaining states, which form 
disordered (chaotic) meshes of points at high energy. 
The regular $L\!=\!0$ states form 
bandheads of $K\!=\!0$ rotational bands. 
Additional $K$-bands corresponding to  
multiple $\beta$ and $\gamma$ vibrations, 
can also be identified. 
An example of such regular $K\!=\!0,2$ bands for 
$\hat{H}_{\mathrm{int}}$ at the critical point, is shown in Fig.~2(a). 
The states in each band have nearly equal values of 
$\langle \hat{n}_d \rangle$, indicating a common intrinsic structure.
\begin{figure}[!t]
\includegraphics[width=0.82\linewidth]{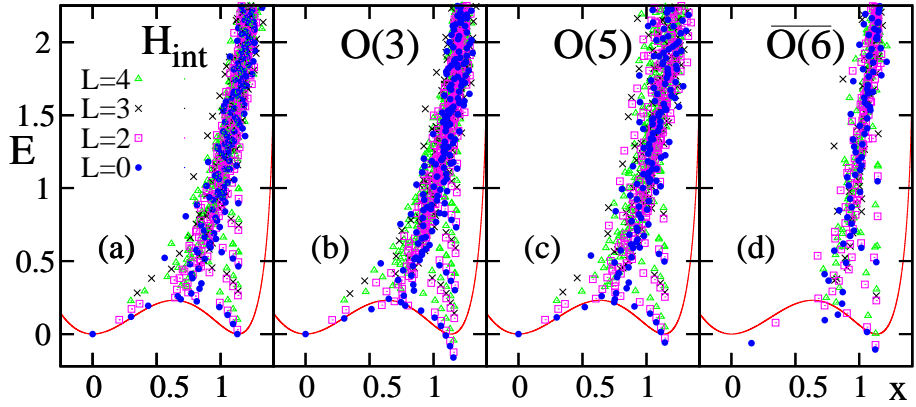}
\caption{
Peres lattices $\{x_i,E_i\}$ for $N\!=\!50,\, L\!=\!0,2,3,4$ 
eigenstates of 
$\hat{H}_\mathrm{int}^{I}(\rho=\rho_c) \!=\! 
\hat{H}_\mathrm{int}^{II}(\xi=\xi_c)$,
Eqs.~(\ref{eq:H1})-(\ref{eq:H2}), 
with $h_2\!=\!1,\,\beta_0\!=\!1.35$ 
[panel (a)] and additional 
collective terms involving $\mathrm{O(3)},\,\mathrm{O(5)}$ 
and $\overline{\mathrm{O(6)}}$ rotations [panels (b), (c), and~(d)]. 
The classical potential shown, is the same in all cases. 
Notice in panels (a)-(b)-(c), 
the well-developed rotational bands 
($K\!=\!0$, $L\!=\!0,2,4$) 
and ($K\!=\!2$, $L\!=\!2,3,4$) formed by the regular states 
in the deformed phase, which are distorted 
in panel~(d).}
\label{fig:2}
\end{figure}

The collective part of the Hamiltonian ($\hat{H}_\mathrm{col}$), 
which does not affect $V(\beta,\gamma)$, 
is composed of the two-body parts of the Casimir operators of the 
groups $\mathrm{O(3),\,O(5)}$ and 
$\overline{\mathrm{O(6)}}$~\cite{ref:Lev87}. 
These orthogonal rotations are associated with the Euler angles, 
$\gamma$ and $\beta$ degrees of freedom, respectively. 
Fig.~\ref{fig:2} shows the Peres lattices corresponding to 
$L\!=\! 0,2,3,4$ eigenstates of 
$\hat{H}_{\mathrm{int}}$ at the critical-point, plus added 
rotational terms one at a time. As seen in Figs.~2(b)-2(c), the 
$\mathrm{O(3)}$ and $\mathrm{O(5)}$ terms preserve the 
ordered $K$-bands of $\hat{H}_{\mathrm{int}}$, Fig.~2(a). 
In contrast, the regular band-structure is strongly disrupted 
by the $\overline{\mathrm{O(6)}}$ term [Fig.~2(d)]. The latter 
couples the deformed  and spherical configurations~\cite{ref:Lev06} 
and mixes strongly the regular and irregular states. 
These results demonstrate the advantage of using the 
resolution $\hat{H} = \hat{H}_{\mathrm{int}} 
+ \hat{H}_{\mathrm{col}}$, since a strong $\overline{\mathrm{O(6)}}$ 
term in the collective part can obscure the simple patterns of the 
dynamics disclosed by the intrinsic part.

This work is supported by the Israel Science Foundation. M.M. acknowledges 
support by the Golda Meir Fellowship Fund and the Czech Ministry of 
Education (MSM 0021620859).


\begin{thebibliography}{12}

\bibitem{ref:Hert76} 
J.A. Hertz, 
\emph{Phys. Rev. B} \textbf{14}, 1165 (1976).

\bibitem{carr} 
L. Carr (Ed.), 
\emph{Understanding Quantum Phase Transitions}, CRC press, 2010.

\bibitem{ref:MacLev11} 
M. Macek, and A. Leviatan, 
\emph{Phys. Rev. C} \textbf{84}, 041302(R) (2011). 

\bibitem{ref:LevMac12} 
A. Leviatan, and M. Macek, 
\emph{Phys. Lett. B} \textbf{714}, 110 (2012).

\bibitem{ref:Cejn10} 
P. Cejnar, J. Jolie, and R.F. Casten, 
\emph{Rev. Mod. Phys.} \textbf{82}, 2155 (2010).     

\bibitem{ref:Iac87} 
F.~Iachello, and A.~Arima,
\emph{The Interacting Boson Model}, 
Cambridge Univ. Press, Cambridge, 1987.

\bibitem{ref:Diep80}
A.E.L.~Dieperink, O. Scholten, and F. Iachello, 
\emph{Phys. Rev. Lett.} \textbf{44}, 1747 (1980).

\bibitem{ref:Whel93} 
N. Whelan, and Y. Alhassid, 
\emph{Nucl. Phys. A} \textbf{556}, 42 (1993).

\bibitem{ref:Lev87} 
A. Leviatan, 
\emph{Ann. Phys. (NY)} \textbf{179}, 201 (1987). 

\bibitem{ref:Heno64} 
M. H\'enon, and C. Heiles, 
\emph{Astron. J.} \textbf{69}, 73 (1964).

\bibitem{ref:Peres84}
A. Peres, 
\emph{Phys. Rev. Lett.} \textbf{53}, 1711 (1984).          

\bibitem{ref:Lev06} 
A. Leviatan, 
\emph{Phys. Rev. C} \textbf{74}, 051301 (2006). 


\end{thebibliography}
\end{document}